\documentstyle[art8,aas2pp4]{article}

\lefthead{Leibundgut et al.}
\righthead{Time dilation in a distant SN~Ia}

\begin{document}

\title{Time Dilation in the Light Curve of the Distant Type~Ia~Supernova 
SN~1995K}

\author{B.~Leibundgut\altaffilmark{1}, R.~Schommer\altaffilmark{2}, 
M.~Phillips\altaffilmark{2}, A.~Riess\altaffilmark{3}, 
B.~Schmidt\altaffilmark{4}, 
J.~Spyromilio\altaffilmark{1}, J.~Walsh\altaffilmark{1}, 
N.~Suntzeff\altaffilmark{2}, M.~Hamuy\altaffilmark{2,5},
J.~Maza\altaffilmark{6}, R.~P.~Kirshner\altaffilmark{3}, 
P.~Challis\altaffilmark{3}, P.~Garnavich\altaffilmark{3}, 
R.~C.~Smith\altaffilmark{7},
A.~Dressler\altaffilmark{8}, R.~Ciardullo\altaffilmark{9}}
\altaffiltext{1}{European Southern Observatory, 
Karl-Schwarzschild-Strasse 2, D--85748 Garching, Germany}
\altaffiltext{2}{Cerro Tololo Inter-American Observatory, 
Casilla 603, La Silla, Chile}
\altaffiltext{3}{Center for Astrophysics, 60 Garden Street, Cambridge,
MA 02138, USA}
\altaffiltext{4}{Mt. Stromlo and Siding Springs Observatories, 
Australian National University, Private Bag,
Weston Creek Post Office, ACT 2611, Australia}
\altaffiltext{5}{Current address: Steward Observatory, University of
Arizona, Tucson, AZ 85721, USA}
\altaffiltext{6}{Departamento de Astronom\'{i}a, Universidad de Chile,
Casilla 36-D, Santiago, Chile}
\altaffiltext{7}{University of Michigan, Department of Astronomy, 834
Dennison, Ann Arbor, MI 48109-1090, USA}
\altaffiltext{8}{Carnegie Observatories, 813
Santa Barbara Street, Pasadena, CA 91101-1292, USA}
\altaffiltext{9}{Pennsylvania State University, Astronomy Department,
525 Davey Laboratory, University Park, PA 16802, USA}

\begin{abstract}
The light curve of a distant Type Ia supernova acts like a clock that
can be used to test the expansion of the Universe.
SN~1995K, at a spectroscopic redshift of $z = 0.479$, provides 
one of the first meaningful data sets for this
test. We find that all aspects of SN~1995K
resemble local supernova Ia events when the light curve is
dilated by $(1+z)$, as prescribed by cosmological expansion. In a
static, non-expanding universe SN~1995K would represent a unique object with a
spectrum identifying it as a regular Type Ia supernova but a light 
curve shape and luminosity which do not follow the well-established relations
for local events. We conclude that SN~1995K provides strong evidence for
an interpretation of cosmological redshifts as due to universal
expansion. Theories in which photons dissipate their energy during travel
are excluded as are age-redshift dependencies.
\end{abstract}


\vspace*{8cm}
\centerline{Accepted for publication in the Astrophysical Journal
Letters}

\section{Introduction}

The nature of galaxy redshifts has usually been
interpreted as due to a general expansion of the universe.
However widely accepted clear experimental proof of this fundamental 
assumption of most cosmological models has been lacking.
The main argument in favor of expansion is the 
observed nearly perfect blackbody energy distribution of the
cosmic background radiation (Mather et al. 1990, Peebles et al. 1991).
The Tolman surface brightness test (Tolman 1930) fundamentally probes
for expansion as well, but implementation of this test has proven
difficult (Sandage \& Perelmutter 1991, Pahre et al. 1996), as
galaxy evolution has to be evaluated independently.
The interpretation of redshifts as due to universal expansion
has been questioned (e.g. Arp 1987, Arp et al. 1990). 
Observations of the apparent clustering 
of high-redshift quasars around
low-redshift galaxies (Burbidge et al. 1971, Arp 1987) and the 
anomalous distribution of redshifts
in groups (Arp 1994) have been used to argue against cosmological
expansion. A theory linking observed redshifts to the ages
of the objects has been developed to explain these findings (Narlikar
\& Arp 1993). 

A direct test of the nature of cosmological redshifts is provided by the
observable effects of time dilation on time variable phenomena at large
redshifts. In an expanding universe these redshifts are directly
related to the change in the scale parameter
inducing a change of distant clock rates for a local observer. The light curve of 
a distant supernova is predicted to be stretched in the observer's frame by a
factor $(1+z)$ compared to the rest frame of the object (Wilson 1939,
Rust 1974, Colgate 1979, 
Tammann 1979, Leibundgut 1990, Hamuy et al. 1993). 
Although the light 
curves of nearby events display, in general, a fairly uniform shape
(Barbon et al. 1973, Leibundgut 1988), recent high-precision
photometry shows that SNe~Ia exhibit differences in their light curves
which are related to their luminosities (Phillips 1993, Hamuy et al. 1995, 
Riess et al. 1995a).

To observe
the real effect one needs a well observed, spectroscopically classified
SN~Ia at a considerable redshift and a thorough 
understanding of the varieties of light curve shapes. 
Attempts to measure the cosmological time dilation with SNe~Ia
have been made before. A sample of nearby events ($z < 0.05$)
was investigated by Rust (1974). For such low redshifts, errors in photometry
and the real variations in light curve shape mask the effect of time dilation. 
The light curve of the distant SN~1988U was employed for a first
test ($z=0.31$; N\o rgaard-Nielsen et al. 
1989, Leibundgut 1991), but the observations
do not cover the maximum and no definite answer could be found.
Recently, the light curve data on other
distant supernovae have been used for a similar analysis
(Goldhaber et al. 1996). 
Here we describe the time dilation test with observations
of SN~1995K, a spectroscopically confirmed SN~Ia with a well-observed
light curve.

\section{The case of SN~1995K}

We have observed a distant supernova superposed on the 
spiral arm of a galaxy at a redshift of 0.479 (Schmidt et al. 1995,
1996). The SN spectrum identifies the object
as a genuine Type Ia supernova displaying the characteristic
[Si~II] absorption around $\lambda_{\rm rest} \approx 6100$\AA\ observed
near 9000\AA and matches that
observed near maximum for local events like SN~1990N or
SN~1989B, but shows distinct differences from
the peculiar examples SN 1991T and SN~1991bg (Schmidt et al. 1996). 

\begin{deluxetable}{l c c c c c c c}
\tablecaption{Fit parameters for $z=0.479$}
\tablehead{
\colhead{Comparison template} 
& \colhead{t$_{\rm max}^B$\tablenotemark{a}}
& \colhead{$\sigma$(t$_{\rm max}^B$)} 
& \colhead{B$_{\rm max}$} 
& \colhead{$\sigma$(B$_{\rm max}$)} 
& \colhead{V$_{\rm max}$} 
& \colhead{$\sigma$(V$_{\rm max}$)} 
& \colhead{$\chi^2$\tablenotemark{b}} }
\startdata
average template & 802.7 & 1.0 & 22.93 & 0.06 & 22.98 & 0.05 & 12.6 \\
SN~1990N         & 801.5 & 1.0 & 22.92 & 0.06 & 22.93 & 0.05 & 12.4 \\
SN~1991T         & 802.5 & 1.1 & 22.94 & 0.06 & 22.96 & 0.05 & 11.1 \\
SN~1992bc        & 799.8 & 1.1 & 22.89 & 0.06 & 22.85 & 0.06 & 12.1 \\
\enddata
\tablenotetext{a}{JD -- 2449000}
\tablenotetext{b}{Degrees of freedom: 21}
\end{deluxetable}

CCD photometry of SN~1995K has been secured in special
filters corresponding to B and V passbands shifted to a redshift of
0.45 and in Kron-Cousins R and I filters. Rest frame B and V light curves
have been constructed from these observations. The close match
of the regular B passband with the R filter at redshift 0.48 reduces the
K-correction terms to a nearly constant value,
independent of the exact color evolution (Kim et al.
1996, Schmidt et al. 1996). The available rest frame B light curve of
SN~1995K covers the supernova peak for over 7 weeks. 
At least one pre-maximum point
from pre-discovery search observations of the field has been recorded.
Since the search was performed with the redshifted B filter we lack
any pre-discovery observations in the rest-frame V passband. 

Simple determination of the decline after maximum indicates a very
slow apparent evolution compared to even the slowest local events (e.g. 
SN~1992bc; Maza et al. 1994, Hamuy et al. 1995). In fact, SN~1995K is the 
slowest
SN~Ia ever observed which we attribute not to intrinsic properties of SN~1995K
but to time dilation.
The photometric observations of SN~1995K provide the basis for this
test of time dilation. The cases of an expanding and a static 
universe are discussed separately in the following sections.
We compare the rest-frame photometry of SN~1995K with light curves of local
supernovae of different decline speeds. As representatives of local
SNe~Ia we have taken SN~1990N (Leibundgut et al. 1991), SN~1991T
(Phillips et al. 1992), SN~1992bc (Maza et al. 1994), and an average
SN~Ia template (Leibundgut 1988). SN~1991T and SN~1992bc
are the slowest SNe~Ia observed to date, while SN~1990N represents a fairly
regular event. All three supernovae have well-observed light curves
from which we constructed comparison templates. SN~1991T
displayed spectral peculiarities near maximum (Filippenko et al. 1992, 
Ruiz-Lapuente et al. 1991, Phillips et al. 1992), which we would have 
detected, if they were present, in SN~1995K. 
The photometry of SN~1995K has been fitted to the light curves
of the local events by means of $\chi^2$ minimization. 
We find that the simplest possible hypothesis fits the facts best: SN~1995K has 
the spectrum and the light curve of a normal SN~Ia, stretched in time by 
a factor (1+0.479).

We could hope to find a close relative of
SN~1995K in the sample of nearby SNe~Ia by determining the best fitting case,
very similar to the procedure employed by Hamuy et al. (1995). The results 
of simultaneous fits to the rest frame B and V photometry are
presented in Table~1. As can be seen, the best fit is achieved for the 
template light curve of SN~1991T. However, no clear trend emerges and all fits are
quite acceptable. The dates and magnitudes at maximum found in the
comparisons agree with each other to
within the errors. This is an encouraging result as it means we
are able to determine these values fairly accurately independently of
the assumptions about the comparison light curve. Conversely, it
indicates the small differences among the comparison templates near
maximum. The variations in $\chi^2$ of the best fits for the different
templates are largely determined by the observations
made well before and well after peak. The sigmas are the
formal errors from the fit and only include the photometric uncertainties 
of the data. Systematic errors certainly increase the uncertainty on the peak
magnitudes.

To demonstrate the robustness of the result, we can estimate the redshift 
of SN~1995K from the photometry 
through the redshift dependence of the light curve stretching.
The results of this analysis is shown in the upper panel of Figure~1.
The deduced redshift changes with the assumed light
curve shape. Interestingly, the formally best fits appear to prefer
a slow supernova at a somewhat smaller redshift than determined from
the spectra, although the differences are small. 
A zero redshift is clearly excluded in all fits.
The light curves of SN~1995K are certainly consistent with the ones of
regular SNe~Ia assuming a time dilation as expected from universal expansion.

To investigate time dilation in a more general manner,
we have performed fits with a parameterized dilation factor of the form
$(1+z)^b$.
However, the data do not constrain the problem enough to fit both parameters
simultaneously. The analysis was performed in such a way that we
determined the global $\chi^2$ of the fit by varying one parameter
while keeping the other at the expected value (0.479 for z and 1 for $b$).
The lower panel
in Figure~1 display the results of this analysis. 
Values of $b>1$ introduce additional light curve
stretching and a compression for $b<1$.
We find that for comparisons with slower
local supernovae, the SN~1995K light curve is too fast for the regular
dilation factor.
The faster light curves reach values closer to 1 as they
need a larger stretching factor to match the SN~1995K photometry. 
Independent fits employing the light curve shape fitting method developed 
by Riess et al. (1995a,b) confirm this result. This 
method is based on a set of known light curve shapes and interpolates 
between them. It further uses a weighing scheme for the model points which 
gives more importance to brighter peak magnitudes and larger uncertainties 
to the fainter parts of the light curve. This is the reason for the higher 
$\chi^2$ of this fit and the much shallower minimum in the $\chi^2$ 
distribution. This method also indicates a solution for 
$b = 1.0^{+0.5}_{-0.25}$ and also clearly excludes small values of $b$.
 
\begin{deluxetable}{l c c c c c c c}
\tablecaption{Fit parameters for non-dilated light curves ($z=0$)}
\tablehead{
\colhead{Comparison template} 
& \colhead{t$_{\rm max}^B$\tablenotemark{a}}
& \colhead{$\sigma$(t$_{\rm max}^B$)} 
& \colhead{B$_{\rm max}$} 
& \colhead{$\sigma$(B$_{\rm max}$)} 
& \colhead{V$_{\rm max}$} 
& \colhead{$\sigma$(V$_{\rm max}$)} 
& \colhead{$\chi^2$\tablenotemark{b}} }
\startdata
average template & 799.1 & 0.5 & 22.27 & 0.06 & 22.54 & 0.05 & 137.7 \\
SN~1990N         & 798.9 & 0.5 & 22.32 & 0.06 & 22.54 & 0.05 & 129.0 \\
SN~1991T         & 799.4 & 0.6 & 22.38 & 0.06 & 22.58 & 0.05 & 90.1 \\
SN~1992bc        & 797.5 & 0.5 & 22.36 & 0.05 & 22.39 & 0.05 & 74.7 \\
\enddata
\tablenotetext{a}{JD -- 2449000} 
\tablenotetext{b}{Degrees of freedom: 21}
\end{deluxetable}

In a static universe time dilation is not expected to act on the light curve.
Redshift in this case is caused by tired light or an equivalent theory
(e.g. the variable mass hypothesis; Narlikar \& Arp 1993)
and is linked to distance
through analyses such as the expanding photospheres in
Type II supernovae (Schmidt et al. 1994) and gravitational lenses 
(Dar 1991). Another
manifestation is the redshift-apparent magnitude diagram of
brightest cluster galaxies (e.g. Postman \& Lauer 1995) and
SNe~Ia. The small scatter in the Hubble diagram of Hamuy et al.
(1995) supports this redshift-distance relation. 
Table~2 lists the fit
parameters for the non-dilated light curve shapes. The global $\chi^2$ 
values clearly exclude these fits. None of the known
light curves of local SNe~Ia is slow enough to match the photometry
of SN~1995K (Figure~2). In particular the maximum magnitude is far from 
the observed
one due to the attempt of the fits to match the pre-maximum point. The
formal errors of the fit parameters are not valid, as can be judged from large
$\chi^2$.

If we take a static universe literally, then SN~1995K was observed at 
an earlier phase (16 days before maximum) than any nearby supernova. 
In that case we are depending on extrapolated pre-maximum points in the 
template light curves, which may not be correct.
Therefore we have removed the pre-maximum point from the SN~1995K photometry
and compared it again with light curves of local SNe~Ia. The quality of
the fits improves dramatically (Figure~2).
The maximum date and magnitude agree much better with
the observations.
Slower light curves are clearly favored in this picture. Nevertheless,
even the slowest local templates are qualitatively worse than
dilated light curves; the evolution of SN~1995K was considerably slower 
than any of the comparison curves.

\section{Discussion}

Figure 2 shows the rest-frame B light curve of SN~1995K 
compared to the best fits of light curves
stretched by the expected factor $(1+z)$ for 
universal expansion and for non-dilated templates.
Two fits for the
non-dilated case are shown which emphasize the importance of the pre-maximum
observation.
The figure demonstrates that without time dilation 
effects, SN~1995K must be a unique event
unrelated to the observational data of local SNe~Ia.
When we assume universal expansion, SN~1995K appears as a
rather normal SN~Ia. The spectrum shows great
similarities to local events which are regarded as non-peculiar,
the color at maximum ($0.0 < B-V < 0.1$) is similar to unreddened nearby
SNe~Ia, indicating little if any absorption, and the luminosity is in 
the range expected from expanding cosmologies (Schmidt et al. 1996). 
The light curve in itself indicates a redshift
which is close to the spectroscopic redshift. 
Complicating the analysis is the variety of light curve shapes 
observed for nearby SNe~Ia. This effect has been interpreted
as an apparent stretching of an underlying basic template
(Perlmutter et al. 1996). However, we know from detailed analysis that the
light curve behavior is more complicated
(Riess et al. 1995a). The data of SN~1995K, unfortunately,
can not distinguish which local supernova provides the best match.
We find the formally best fits to indicate a slightly lower redshift or,
equivalently, a slightly retarded cosmological expansion. All fits are
determined very strongly by the pre-maximum observation and the latest
data points.
This highlights the importance for extended coverage of
SN~Ia events to perform this time dilation test. In addition, the photometric
accuracy of the data critically determines the goodness of the fits.

In a static universe the Hubble constant is time-independent
and just measures the red\-shift-distance proportionality. 
For a conventional Hubble constant of $H_0 = 50$~km~s$^{-1}$ Mpc$^{-1}$ 
one unit in
redshift corresponds to 6000~Mpc. The same number for $H_0 =
80$~km~s$^{-1}$~Mpc$^{-1}$ is 3750~Mpc. The
luminosity of SN~1995K in such a static universe is
$M_B=-19.3 + 5 \log(H/50)$.
Our best
estimate for the absolute magnitude SN~1995K should have when we use the 
decline rate
relation of Hamuy et al. (1995), however, is $M_B=-20.4\pm0.2+5\log(H/50)$, 
with $\Delta m_{15} \approx 0.5$ and the most conservative 
estimate of the decline-luminosity relation (eq. 11 of Hamuy et al. 1995). 
This means SN~1995K should be about 1
magnitude more luminous than what would be observed in a static universe model.
Note that the extrapolation goes well beyond the set of
objects on which the method is based ($0.8 < \Delta m_{15} < 1.5$). 
Even compared to the average absolute magnitude of local supernovae ($M_B =
-19.7\pm0.25+5\log(H/50)$) SN~1995K appears underluminous.
In a static universe, SN~1995K would have been a truly unique SN~Ia -
exhibiting the slowest known photometric evolution, yet displaying a
normal spectrum, and being substantially less luminous than its nearby
counterparts.

The apparent peak magnitude of SN~1995K can also be compared to the
expectation of a steady-state model. Since the
universe is expanding in this model, time dilation cannot distinguish
it from the standard Big Bang. However, the steady-state model
predicts a luminosity-redshift relation. With the sample of nearby
SNe~Ia compiled by Hamuy et al. (1995) we can predict the expected
apparent magnitude of a distant supernova given its redshift.
With equation (9) from Hamuy et al., i.e. excluding any luminosity
corrections from the light curve shape, we predict a peak magnitude of
SN~1995K of B=$23.4 \pm 0.1$ where the error is composed from the
uncertainty in the determination of the zero-point (cf. difference
between eqs. (3) and (9) in Hamuy et al.). The scatter of the local
supernovae around the relation is $\sigma = 0.25$. SN~1995K was observed
about 0.4~magnitude brighter than this prediction. With this single
measurement we are thus not able to exclude the steady-state model due
to the intrinsic scatter in the luminosity of SNe~Ia.

\section{Conclusions}

The photometry of SN~1995K extending over about 50 days provides
sufficient data to probe the effect of time dilation on a clock
running at a cosmological distance. We find that including the 
time dilation
expected from universal expansion makes SN~1995K comparable to local SNe~Ia
and a fair representative of its class. 
The spectrum, color, luminosity at maximum, and the light curve shape
are all very similar to what is observed in local Type Ia supernova. 
Together with
more SNe~Ia at cosmological distances it can be used to determine the
deceleration parameter and contributions of non-baryonic mass to the 
cosmic mass density.

On the other hand, assuming a static universe, SN~1995K had the
slowest photometric evolution of all known SNe~Ia despite appearing
spectroscopically indistinguishable from local events. The photometry of
SN~1995K cannot be approximated by any light curve shape of
nearby events and, in a non-expanding universe, does also not follow the
decline-luminosity relation established for nearby events. It is even
less luminous than the mean of local supernovae and would constitute a unique and
peculiar SN~Ia.

We take this as
a clear vindication of an expanding universe. Further
supernovae at redshifts beyond 0.1 provide additional checks (cf.
Goldhaber et al. 1996).
Such distant objects are currently discovered at a regular rate
(Perlmutter et al. 1995a,b, 1996, Kirshner et al.
1995, Garnavich et al. 1996a,b) and will provide
additional tests for consistency. The importance of extensive light
curve coverage as early and as long as possible must be stressed. It
is only these observations which provide enough leverage to perform
this test. 
The early and
late phase photometry is also of paramount importance for an accurate
determination of the deceleration parameter $q_0$ in order to identify
the best local counterpart and find the most appropriate luminosity.
Direct observation of the detailed spectral evolution
of a distant supernova provides a further test. Here the changes in
line shifts and relative strengths would reveal the apparently retarded
evolution of the supernova.

\section{References}

\begin{list}{}%
{\setlength {\itemindent -10mm} \setlength {\itemsep 0mm} \setlength%
{\parsep 0mm} \setlength {\topsep 0mm}}
\item Arp, H. C. 1987, Quasars, Redshifts and Controversies, Interstellar Media, Berkeley
\item Arp, H. C. 1994, ApJ, 430, 74
\item Arp, H. C., Burbidge, G. R., Hoyle, F., Narlikar, J. V., \& Wickramasinghe, N. C. 1990, Nature, 346, 807
\item Barbon, R., Ciatti, F., \& Rosino, L. 1973, A\&A, 25, 241
\item Burbidge, E. M., Burbidge, G. R., Solomon, P. M., \& Strittmatter, P. A. 1971, ApJ, 170, 233
\item Colgate, S. A. 1979, ApJ, 232, 404
\item Dar, A. 1991, ApJ, 382, L1
\item Filippenko, A. V., et al. 1992, ApJ, 384, L15
\item Garnavich,, P., et al. 1996a, IAU Circ. 6332
\item Garnavich,, P., et al. 1996b, IAU Circ. 6358
\item Goldhaber, G., et al. 1996, Thermonuclear Supernovae, eds. R. Canal, P. Ruiz-Lapuente, \& J. Isern, (Dordrecht: Kluwer), in press
\item Hamuy, M., Phillips, M. M., Wells, L. A., \& Maza, J. 1993, PASP, 105, 787
\item Hamuy, M., Phillips, M. M., Maza, J., Suntzeff, N. B., Schommer, R. A., \& Avil\'es, R. 1995, AJ, 109, 1
\item Kim, A., Goobar, A., \& Perlmutter, S. 1996, PASP, in press
\item Kirshner, R. P., et al. 1995, IAU Circ. 6267
\item Leibundgut, B. 1988, Ph.D. Thesis, University of Basel
\item Leibundgut, B. 1990, A\&A, 229, 1
\item Leibundgut, B. 1991, Supernovae, ed. S. E. Woosley, (New York: Springer), 751
\item Leibundgut, B., et al. 1991, ApJ, 371, L23
\item Mather, J., et al. 1990, ApJ, 354, L37
\item Maza, J., Hamuy, M., Phillips, M. M., Suntzeff, N. B., Avil\'es, R. 1994, ApJ, 424, L107
\item Narlikar, J. \& Arp, H. 1993, ApJ, 405, 51
\item N\o rgaard-Nielsen, H. U., Hansen, L., J\o rgensen, H. E., Salamanca, A. A., Ellis, R. S., \& Couch, W. J. 1989, Nature, 339, 523
\item Pahre, M. A., Djorgovski, S. G., \& de Carvalho., R. R. 1996, ApJ, 456, L79
\item Peebles, P. J. E., Schramm, D. N., Turner, E. L., \& Kron, R. G. 1991, Nature, 352, 769
\item Perlmutter, S., et al. 1995a, ApJ, 440, L41
\item Perlmutter, S., et al. 1995b, IAU Circ. 6270
\item Perlmutter, S., et al. 1996, Thermonuclear Supernovae, eds. R. Canal, P. Ruiz-Lapuente, \& J. Isern, (Dordrecht: Kluwer), in press
\item Phillips, M. M. 1993, ApJ, 413, L105
\item Phillips, M. M., et al. 1992, AJ, 103, 1632
\item Postman, M. \& Lauer, T. R. 1995, ApJ, 440, 28
\item Riess, A. G., Press, W. M., \& Kirshner, R. P. 1995a, ApJ, 438, L17
\item Riess, A. G., Press, W. M., \& Kirshner, R. P. 1995b, ApJ, 445, L91
\item Ruiz-Lapuente, P., Cappellaro, E., Turatto, M., Gouiffes, C. Danziger, I. J., Della Valle, M., \& Lucy, L. B. 1992, ApJ, 387, L33
\item Rust, B. W. 1974, Ph.D. Thesis, Oak Ridge National Laboratory (ORNL-4953)
\item Sandage, A. \& Perelmutter, J.-M. 1991, ApJ, 370, 455
\item Schmidt, B. P., Kirshner, R. P., Eastman, R. G., Phillips, M. M., Suntzeff, N. B., Hamuy, M., Maza, J., \& Avil\'es, R. 1994, ApJ, 432, 42
\item Schmidt, B. P., et al. 1995, IAU Circ. 6160
\item Schmidt, B. P., et al. 1996, in preparation
\item Tammann, G. A. 1978, Astronomical Uses of the Space Telescope, eds. F. Macchetto, F. Pacini \& M. Tarenghi (Garching: ESO Proceedings), 329
\item Tolman, R. C. 1930, Proc. Nat. Acad. Sci., 16, 511
\item Wilson, O. C. 1939, ApJ, 90, 634
\end{list}



\end{document}